\documentclass[final,5p,times,twocolumn,authoryear]{elsarticle}
\usepackage{amsmath,amssymb}
\usepackage{mathrsfs}
\usepackage{graphicx}
\usepackage{color}
\usepackage{setspace}
\usepackage{lscape}
\usepackage{threeparttable}
\RequirePackage[l2tabu,orthodox]{nag}
\usepackage[all,warning]{onlyamsmath}
\newtheorem{ass}{Assumption}
\newtheorem{prop}{Proposition}

\newcommand{\II}{I\hspace{-0.1em}I }
\newcommand{\III}{I\hspace{-0.1em}I\hspace{-0.1em}I }
\newcommand{\piBEL}{\pi^{B\!E\!L}}
\newcommand{\piBEU}{\pi^{B\!E\!U}}
\newcommand{\piSEL}{\pi^{S\!E\!L}}
\newcommand{\piSEU}{\pi^{S\!E\!U}}
\newcommand{\thetaSEL}{\theta^{S\!E\!L}}

\newcommand{\thetaBSL}{\theta^{B\!S\!/L}}
\newcommand{\thetaBSU}{\theta^{B\!S\!/U}}
\newcommand{\thetaSE}{\theta^{S\!/\!E}}
\newcommand{\thetaSn}{\theta^{S\!-}}
\newcommand{\thetaSp}{\theta^{S\!+}}
\newcommand{\piS}{\pi^{S}}
\newcommand{\piE}{\pi^{E}}
\begin{document}
\begin{frontmatter}
\title{Price and Fulfillment Strategies in Omnichannel Retailing}
\author{Yasuyuki Kusuda\corref{cor1}}
\cortext[cor1]{Associate Professor.\ead{kusuda@n-fukushi.ac.jp}}
\address{Faculty of Economics, Nihon Fukushi University, Aichi 477-0031, Japan}
\begin{abstract}
Omnichannel retailing, a new form of distribution system, seamlessly integrates the Internet and physical stores. This study considers the pricing and fulfillment strategies of a retailer that has two sales channels: online and one physical store. The retailer offers consumers three purchasing options: delivery from the fulfillment center, buy online and pick up in-store (BOPS), and purchasing at the store.
Consumers choose one of these options to maximize their utility, dividing them into several segments.
Given the retailer can induce consumers to the profitable segment by adjusting the online and store prices, our analysis shows that it has three optimal strategies:
(1) The retailer excludes consumers far from the physical store from the market and lets the others choose BOPS or purchasing at the store.
(2) It lets consumers far from the physical store choose delivery from the fulfillment center and the others choose BOPS or purchasing at the store.
(3) It lets all consumers choose delivery from the fulfillment center.
Finally, we present simple dynamic simulations that considers how the retailer's optimal strategy changes as consumers' subjective probability of believing the product is in stock decreases.
The results show that the retailer should offer BOPS in later periods of the selling season to maximize its profit as the subjective probability decreases.
\end{abstract}
\begin{keyword}
Omnichannel \sep Fulfillment \sep Consumer behavior
\end{keyword}
\end{frontmatter}
\section{Introduction}
Omnichannel retailing, a new form of distribution system, seamlessly integrates the Internet and physical stores.
From consumers' point of view, products ordered online can be collected at a nearby physical store (termed ``buy online and pick up in-store,'' or BOPS, herein), thus saving the shipping cost borne by consumers.
On the contrary, from the retailer's point of view, to make such new sales options available to consumers, it is necessary to reconstruct the flow of fulfillment for the sale of products as a new distribution system.
For example, for products ordered online, the retailer can deliver them from the fulfillment center or allow the consumer to collect them at a physical store.
In addition, if the retailer assigns a certain amount of inventory to each physical store in advance, some of the fulfillment of products ordered online to the physical store can be left.
However, these new attributes of the distribution system have raised a number of questions such as what fulfillment should be used when cross-channel fulfillment is available, when should it be adopted, and what optimal pricing strategy is consistent with such fulfillment?
\par
\citet{Lei_Jasin_Sinha2018} is the first study to describe the price and fulfillment problems retailers must solve in a distribution system comprising multiple sales regions and distribution centers.
They compare two optimization models.
The first model is a two-stage model in which, in the first stage, the retailer optimizes the price under the constraint that total demand in the regions does not exceed the total inventory of the fulfillment centers. In the second stage, it optimizes fulfillment from the centers to regions at the optimal price.
In the second model, termed the joint pricing and fulfillment problem, the retailer optimizes the optimal price and fulfillment simultaneously.
Comparing these two models, they point out that the latter can earn the retailer a higher profit than the former.
Clearly, the intuitive implication for the profitability of the second model is that the retailer can minimize fulfillment costs by raising the price for the region with costly shipping.
Lei et al. (2018) consider a multi-period joint pricing and fulfillment problem in which the retailer sells multiple products in different demand regions by changing prices to maximize total expected profits.
\citet{Harsha_Subramanian_Uichanco2019}, on the contrary, apply the idea of \citet{Lei_Jasin_Sinha2018} to an omnichannel distribution system consisting of one fulfillment center and multiple regions (\emph{zones}) with physical stores in which product orders placed online can be fulfilled by either the fulfillment center or a physical store.
However, they assume that the fulfillment of products ordered online is random and exogenous because it may occur through BOPS at consumers' request.
\citet{Harsha_Subramanian_Uichanco2019}
thus propose heuristics for this multi-period problem as the centralized optimization problem with endogenous fulfillment by introducing the idea of the ``partition'' of inventory.
\par
These previous studies suffer from two main problems.
First, fulfillment is often assumed to be random and exogenous because it cannot be fully coordinated\footnote{Dynamic pricing models often use the assumption of Poisson arrival.
See, for example, \citet{Gallego_van-Ryzin1997}.}.
According to
\citet{Harsha_Subramanian_Uichanco2019}, in practice, fulfillment may be managed by decentralized order management systems with complicated rules that depend on store performance, fulfillment capacity, and so on\footnote{Indeed, their study collaborated with IBM Commerce.}.
However, omnichannel distribution is, by definition, a cross-channel system that uses fulfillment centers and a network of physical stores.
An omnichannel retailer (or the headquarters) might thus be able to manage fulfillment for multiple regions centrally, at least to some extent, by coordinating channels using, say, a central information system. The second problem with previous studies is that their models do not fully reflect consumers' behavior in terms of choosing BOPS for products ordered online\footnote{Instead, some studies simply assume consumers' channel demand in a multinomial logit probability model.
For example, see Equation (7) in \citet{Harsha_Subramanian_Uichanco2019}.}.
If the models include consumer behavior, BOPS demand cannot be a random variable but rather must be the optimal solutions for consumers who minimize their disutility in terms of receiving the ordered products.
\par
These two problems motivate us to propose a new framework that includes both price and fulfillment decisions in a centralized scheme with endogenous consumer behavior. Specifically, to address the shortcomings of previous research, this study considers the pricing and fulfillment strategies of a retailer that has two sales channels: online and one physical store.
The consumers that purchase a product from this retailer are assumed to be located different distances from the physical store and thus incur different travel costs. Meanwhile, the retailer offers consumers three purchasing options: delivery from the fulfillment center, BOPS, and purchasing at the store.
Consumers choose one of these options to maximize their utility, dividing them into several segments.
Based on the foregoing, the research questions of this study are as follows:
\begin{enumerate}
\item Can the retailer lead consumers to the profitable segment by adjusting its online price and store price?
\item If possible, what kinds of segments are formed?
\item How much profit can be obtained from the pattern of these segments?
\item What are the optimal prices?
\end{enumerate}
There are two critical points here: the difference between the two prices and the subjective probability that the consumer will believe that the product is in stock at the physical store (simply termed the subjective probability hereafter).
If the online price is sufficiently higher than the store price, the consumer is induced to visit the physical store.
Meanwhile, if the subjective probability is low, the consumer is induced to shop online.
On the contrary, the retailer must bear the costs of the three fulfillment options (i.e., its fulfillment costs).
Therefore, the problem is how the retailer that observes the subjective probability optimizes its profit considering its fulfillment costs.
\par
To address these problems, this study clarifies the following points.
First, the retailer can induce consumers to the profitable segment by adjusting the online and store prices.
By controlling the number of consumers who choose delivery from the fulfillment center, BOPS, or purchasing at the store, the retailer can generate additional profit.
Consumer segments are formed by dividing consumers by proximity to the physical store, revealing how many in each segment are willing to purchase the product.
This study clarifies the segment patterns formed by the two prices and shows how much profit the retailer can earn from each pattern.
Next, we find the retailer's optimal strategies given those profits based on the two optimal prices and the optimal consumer segments.
Our analysis shows that the retailer has three optimal strategies.
In the first, it excludes consumers far from the physical store from the market and lets those close to it choose BOPS or purchasing at the store.
The second optimal strategy lets consumers far from the physical store choose delivery from the fulfillment center and those close to it choose BOPS or purchasing at the store.
In the final strategy, all consumers choose delivery from the fulfillment center.
The optimal strategy of these depends on the fulfillment costs associated with delivery from the fulfillment center, BOPS, and purchasing at the store.
If the cost for BOPS is sufficiently higher than the cost of selling at the store, the retailer must let consumers close to the physical store choose to purchase at the store and BOPS otherwise.
We also find that the retailer lets all consumers choose delivery if that option's cost is sufficiently lower than those of the other two options.
Finally, we present simple dynamic simulations that considers how the retailer's optimal strategy changes as consumers' subjective probability decreases.
In the stage in which the subjective probability is high, consumers choose to purchase at the store.
When this probability decreases, the retailer switches its strategy at a certain point to allow them to choose BOPS.
However, if the store's inventory runs out during the sales period, this leads all consumers to choose delivery.
\par
This study extends research on the joint pricing and fulfillment problem as well as on the subjective probability, such as \citet{Gao_Su2016} and \citet{Kusuda2019}.
Those studies assume a ``rational expectation'' of the consumer's subjective probability, on which the optimal inventory problem depends; in turn, the optimal strategy is consistent with the formation of the consumer's subjective probability. In contrast to that body of the literature, this study rather assumes a completely exogenous subjective probability and considers simple dynamic simulations in which it decreases over time.
\par
This study contributes to the research on this topic by extending the traditional dynamic pricing model (also called the yield management model).
Dynamic pricing is a price discrimination model in which retailers sell their unreplenished inventory (or \emph{resource}) of goods or services (or \emph{plan}) such as air tickets and hotel rooms using several options and prices in a selling period\footnote{For a survey of dynamic pricing models, see \citet{Bitran_Caldentey2003}, \citet{Elmaghraby_Keskinocak2003}, \citet{Talluri2006}, and \citet{Chen_Simchi-Levi2012}.}.
In recent years, such models have been applied to the field of e-commerce \citep{Shpanya2014}.
In the dynamic pricing model, retailers sell to consumers at several prices by dynamically allocating resources to several plans.
Meanwhile, the omnichannel model can be regarded as an allocation model in which retailers use inventory in fulfillment centers and physical stores for fulfillment.
In the simple example of aircraft tickets provided by \citet{Talluri2006}, taking advantage of differences in passenger preferences, retailers can sell tickets with a limited option at a low price to tourist passengers and those with an unlimited option at a high price to business passengers.
Similarly, in this omnichannel model, the retailer takes advantage of differences in consumers' disutility when they visit the physical store.
It can then sell the store inventory to consumers living close to the physical store and that in the fulfillment center to consumers living far from it.
Furthermore, if consumers' subjective probability changes dynamically, the retailer can also dynamically change the sales strategy by offering several fulfillment options.
This study therefore provides a solution to how retailers set prices dynamically in practice, which offers another perspective on the dynamic pricing literature\footnote{However, this study's dynamic simulation is a ``myopic'' one in which the retailer observes only the current subjective probability to decide the current price, which is not a formal dynamic model.}.
\par
The remainder of the paper is organized as follows. Section 2 describes the model used for the analysis.
Section 3 considers the retailer's strategies based on that model.
In Section 4, the results of simple dynamic simulations are presented. Section 5 concludes.
\section{The model}
\subsection{Three options and consumer behavior}
In this model, one retailer has two sales channels: online and one physical store. The retailer sells a simple product through both sales channels.
Here, physical store sales occur when consumers directly visit stores to purchase products and online sales occur when they order from the retailer's website. Its website can also be used to communicate with consumers (e.g., to indicate the availability of the BOPS service)\footnote{See \citet{Kusuda2019} for the actual BOPS systems implemented by online retailers such as Amazon.}.
Consumers who order products online can choose between delivery from the fulfillment center to their home and BOPS (if available).
If a consumer chooses BOPS, the retailer uses the store inventory as long as there is stock.
The retailer can set different prices for online sales and store sales and consumers can observe these two prices online in advance.
The online price is the same for both delivery and store purchase.
After observing the physical store's distance, subjective probability, and price by sales channel, consumers choose one of the three options: delivery, BOPS, and store purchase.
As a result, demand for each option is determined by consumers' preference.
\par
The utility of consumers when choosing one of these three options is formulated as follows.
First, let the shipping cost per unit when the retailer delivers from the fulfillment center be a uniform constant, $c\ (> 0)$. We make the following assumption about this shipping cost.
\begin{ass}
Shipping cost $c$ is borne by the consumer, not the retailer.
\end{ass}
This assumption is customary and reasonable given the postage and delivery charges of third-party companies.
\par
\begin{figure}[htbp]
\includegraphics[width=8.5cm,bb=0 0 297.75 141.75]{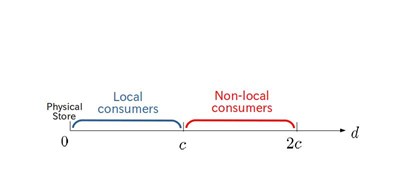}
\caption{Consumer space}
\label{fig:Segment_space}
\end{figure}

Next, let $v\ (>c)$ be the value obtained from this product (common to all consumers) and assume that potential consumers who may purchase the product are distributed in the interval $[0, v]$ at a density of $1$ with the physical store existing at point $0$.
In other words, the distance $d \in [0, v]$ of a consumer from the physical store can be referred to as the \emph{travel cost} that the consumer bears to visit that store.
Thus, potential consumers farthest from the physical store must be marginal consumers who would visit it even at price 0, and we exclude consumers at $d > v$ from the model.
Let us call the consumers located at $d \in [0, c]$ \emph{local consumers} and those at $d \in (c, v]$ \emph{non-local consumers} for convenience.
Note that $d$ is the cost incurred by consumers for the BOPS option and $c$ is the delivery cost.
Then, if all consumers purchase online, all local consumers choose BOP and all non-local consumers choose delivery.
Therefore, to make the discussion more convenient, let us assume that the number of local and non-local consumers is equal\footnote{The number 2 in this assumption is not essential to the discussion; it is just made for simplicity.
In other words, even assuming $v = kc \ (k > 0)$ does not lose generality.}.
\begin{ass}
The value of a product is equal to twice the shipping cost, $v = 2c$.
\end{ass}
This assumption makes boundary consumers ($d = c$) indifferent between delivery (paying $c$) and BOPS (paying $d$).
Figure \ref{fig:Segment_space} shows the consumer regions.
In the following analysis, we assume that each consumer in the interval $[0, 2c]$ purchases at most one unit of the product.
\par
%
%
%
\par
Even when visiting the physical store directly without ordering online, the consumer cannot purchase the product if the physical store is out of stock.
In this model, if a consumer cannot purchase the product in the physical store, he/she cannot do so in other ways on that day\footnote{This assumption is merely made for simplicity.
In reality, such consumers could order online the same day after returning home, but this scenario complicates the analysis.}.
Then, consumers visiting the physical store without ordering online predict whether the product is in stock in the store at the subjective probability, $\theta \in [0, 1]$, which is identical for all consumers and common knowledge between them and the retailer.
Furthermore, we assume that this probability is exogenous to the retailer and uncorrelated with the physical store's actual inventory\footnote{\citet{Gao_Su2016} and \citet{Kusuda2019} analyze the subjective probability of such consumers under the assumption of rational expectations.
By contrast, in this analysis, the subjective probability is entirely exogenous to the retailer, while it is common knowledge to it and the consumers.
This assumption may seem unnatural, but considering $\theta$ as a parameter makes the following analysis more manageable.}.
\par
Let $p_{e}$ and $p_{s}$ denote the online price and store price, respectively.
Then, the respective (expected) utilities of consumers choosing delivery, BOPS, and store purchase are
\begin{equation}
\label{eq:utility}
\begin{aligned}
u_{e} & \equiv v - p_{e} - c = c - p_{e}, \\
u_{b} & \equiv v - p_{e} - d = 2c - p_{e} - d,
\ \ d \in [0, 2c], \\
u_{s} & \equiv \theta(v - p_{s}) - d = 2\theta c - \theta p_{s} - d,
\ \ d \in [0, 2c].
\end{aligned}
\end{equation}
Here, consumers incur the shipping cost when choosing delivery and the travel cost when choosing BOPS.
If choosing the store purchase option, they must incur the travel cost and take the risk the product is out of stock\footnote{In this utility setting, the only costs (disutility) other than the price borne by the consumer are $c$, $d$, and $\theta$.
Here, we might consider another disutility due to ``waiting time'' when ordering online.
However, we ignore this problem by including such disutility in the shipping cost, $c$.}.
While utility with delivery is common to all consumers, utility with BOPS or store purchase varies depending on each consumer's travel cost, $d$.
Further, assume that consumers are risk-neutral and that their reservation utility is $0$.
Therefore, we set the following assumption.
\begin{ass}
Each consumer purchases the product as long as $u_{i} \geq 0$ and does not otherwise ($i = e, b, s$).
\end{ass}
\subsection{Consumer segments}
Next, we divide the consumers distributed in the interval $[0, 2c]$ into three segments according to their behavior.
Let $\mathscr{S}_{i}, \ i = e, b, s$ denote the segment for consumers with $u_{i} \geq \max\{ u_{j}, u_{k} \}, \ i = e, b, s; \ j, k \neq i$.
Consumers can then be classified as follows based on their travel cost $d \in [0, 2c]$:
\begin{equation*}
\label{eq:segment}
\begin{aligned}
	d \in \mathscr{S}_{e}: \ \ &
	  d > c \ \ \text{and} \ \ \textcolor{black}{\delta \leq (1 - 2\theta)c + d}, \\
	d \in \mathscr{S}_{b}: \ \ &
	  d \leq c \ \ \text{and} \ \ \delta \leq (2 - 2\theta)c, \\
	d \in \mathscr{S}_{s}: \ \ &\
	  \delta > (2 - 2\theta)c \ \ \text{and} \ \ \textcolor{black}{\delta > (1 - 2\theta)c + d},
\end{aligned}
\end{equation*}
where $\delta \equiv p_{e} - \theta p_{s}$.
The size of $\theta$ compared with $\delta$ and $d$ determines the segment to which a consumer belongs.
We classify three cases: \emph{Stage I} for $1/2 < \theta \leq 1$, \emph{Stage II} for $0 < \theta \leq 1/2$, and \emph{Stage III} for $\theta = 0$\footnote{We use the word ``stage'' as it implicitly assumes that the subjective probability, $\theta$, decreases over time.}.
However, in this model, because each segment is defined only by the price difference, $\delta$, no unique solution for the optimal prices exists.
Therefore, the following assumption is necessary to find the unique solution.
\begin{ass}
In all the cases, the retailer cannot set a price that exceeds the value of the product: $p_{e}, p_{s} \leq 2c \ \textcolor{black}{(= v)}$.
\end{ass}
Although this assumption is only for the solution to be definitive, it would be natural to set the product's value as the price's upper limit\footnote{Furthermore, as seen later, the consumer segment's pattern (and demand) depends on the size of $\delta$.
Therefore, without this assumption, there exists an infinite number of prices set $(p_{e}, p_{s})$ for the same $\delta$, which leaves the solution undetermined.}.
Note that $-2\theta c \leq \delta \leq 2c$ from this assumption.
\par
\begin{figure}[htbp]
\centering
\includegraphics[width=5.5cm,bb=0 0 297.75 297.75]{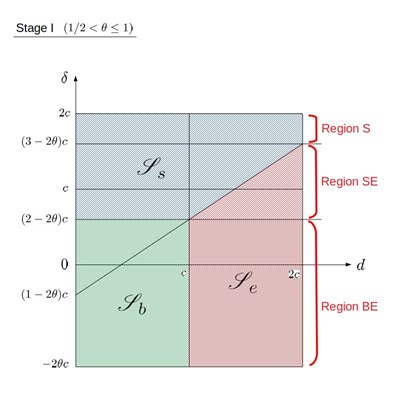}
\caption{Consumer segments (Stage I with BOPS)}
\label{fig:Segment_F1_BOPS}
\end{figure}
\begin{figure}[htbp]
\centering
\includegraphics[width=5.5cm,bb=0 0 297.75 297.75]{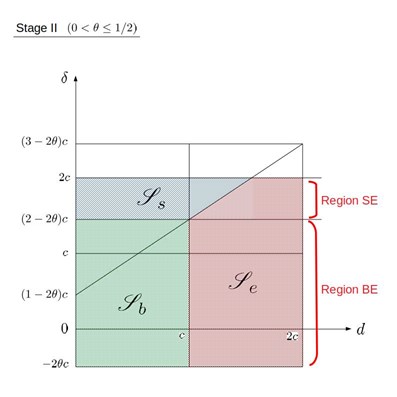}
\caption{Consumer segments (Stage \II with BOPS)}
\label{fig:Segment_F2_BOPS}
\end{figure}
\begin{figure}[htbp]
\centering
\includegraphics[width=5.5cm,bb=0 0 297.75 297.75]{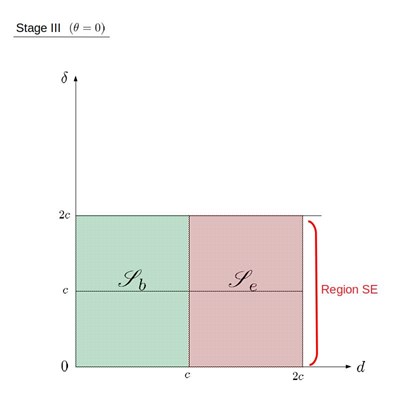}
\caption{Consumer segments (Stage \III with BOPS)}
\label{fig:Segment_F3_BOPS}
\end{figure}

Figures \ref{fig:Segment_F1_BOPS}--\ref{fig:Segment_F3_BOPS} show the consumer segments by classifying them on the $(d\!-\!\theta)$ plane.
As shown, when $\delta$ is low, delivery and BOPS are advantageous for consumers; by contrast, when $\delta$ is high, delivery and store purchase are beneficial.
Furthermore, in Stage I, as $\delta$ approaches $2c$, all consumers choose store purchase.
\par
%
%
\noindent
Now, let us classify the consumer segment patterns as follows:
\begin{equation*}
\label{eq:region1}
\begin{aligned}
	& \textbf{Region BE}\ (-2 \theta_{e} \leq \delta \leq (2 - 2\theta)c): \\
  & \ \ d \in [0, c] \Rightarrow d \in \mathscr{S}_{b}; \
    d \in (c, 2c] \Rightarrow d \in \mathscr{S}_{e} \\
	& \textbf{Region SE}\ ((2 - 2\theta)c < \delta \\
  & \hspace{3cm} \leq \min\{ (3 - 2\theta)c, 2c \}): \\
  & \ \ d \in [0, \delta - (1 - 2\theta)c] \Rightarrow d \in \mathscr{S}_{s}; \\
  & \ \ d \in (\delta - (1 - 2\theta)c, 2c] \Rightarrow d \in \mathscr{S}_{e} \\
	& \textbf{Region S}\ ((3 - 2\theta)c < \delta \leq 2c, \text{only Stage I}): \\
  & \ \ d \in \mathscr{S}_{s} \ \text{for all $d$.}
\end{aligned}
\end{equation*}
In other words, Region BE is divided into delivery and BOPS, Region SE is divided into delivery and store purchase, and Region S is classified as store purchase for all consumers.
\subsection{Demand and profits}
Next, we formulate the demand function of the retailer and its profit maximization.
First, the retailer bears the fulfillment costs.
Let $c_{e}$, $c_{b}$, and $c_{s}$ denote the fulfillment costs for delivery, BOPS, and store purchase, respectively, assuming all the variables are non-negative.
These costs are exogenous variables representing employees' fulfillment efforts at the fulfillment center and physical store.
The fulfillment cost for delivery, $c_{e}$, does not include the shipping cost, $c$.
In addition, we do not consider the inventory carrying costs in this model\footnote{This is because this analysis focuses only on the fulfillment cost problem. Moreover, we ignore the inventory amount in the fulfillment center and physical store in this analysis.}.
\par
\begin{figure}[htbp]
\centering
\includegraphics[width=8.5cm,bb=0 0 297.75 297.75]{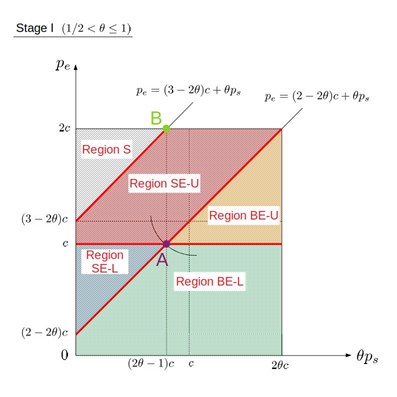}
\caption{Price regions (Stage I with BOPS)}
\label{fig:Region_F1_BOPS}
\end{figure}
\begin{figure}[htbp]
\centering
\includegraphics[width=8.5cm,bb=0 0 297.75 297.75]{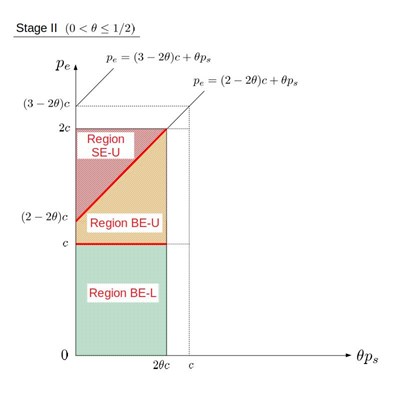}
\caption{Price regions (Stage \II with BOPS)}
\label{fig:Region_F2_BOPS}
\end{figure}

Here, we find the demand and profit function for each region and maximize the profit to have the optimal price.
First, because the consumer segment merely refers to the subset of consumers who choose one of the three options, not all of the consumers in each segment necessarily make a purchase.
Indeed, as shown in Assumption 3, consumers do not purchase unless their utility is no less than the reservation level.
In particular, if $p_{e} > c$, no consumers---even in Regions BE and SE---choose delivery because $u_{e} < 0$.
Therefore, we divide Regions BE and SE into two to classify them as follows:
\begin{equation*}
\begin{aligned}
\label{eq:region2}
  \text{In Region BE};\ \
  p_{e} \leq c \Rightarrow \textbf{Region BE-L}, \\
  p_{e} > c \Rightarrow \textbf{Region BE-U}. \\
  \text{In Region SE};\ \
  p_{e} \leq c \Rightarrow \textbf{Region SE-L}, \\
  p_{e} > c \Rightarrow \textbf{Region SE-U}.
\end{aligned}
\end{equation*}
Figures \ref{fig:Region_F1_BOPS} and \ref{fig:Region_F2_BOPS} illustrate each region on the $(p_{e}\!-\!\theta p_{s})$ plane.
As shown in these figures, the price plane is divided into five regions (BE-L, BE-U, SE-L, SE-U, and S) in Stage I, three regions (BE-L, BE-U, and SE-U) in Stage II, and two regions (BE-L and BE-U) in Stage III (figure omitted).
\par
\vspace{\baselineskip}
\vspace{\baselineskip}
\noindent
Because product demand is equal to the number of consumers who purchase, we can find the demand function for each region, as the following proposition shows. (The proof is shown in Appendix A.)
\begin{prop}
\label{prop:demand}
Demand for the product in each region is as follows:
$D_{e} = c,\ D_{b} = c$ for Region BE-L;
$D_{e} = 0,\ D_{b} = 2c - p_{e}$ for Region BE-U;
$D_{e} = (3 - 2\theta)c - p_{e} + \theta p_{s},\ D_{s} = -(1 - 2\theta)c - \theta p_{s} + p_{e}$ for Region SE-L;
$D_{e} = 0,\ D_{s} = 2\theta c - \theta p_{s}$ for Region SE-U;
and
$D_{s} = 2\theta c - \theta p_{s}$ for Region S.
\end{prop}
The retailer has two demand options.
One option expects demand from both delivery and BOPS or both delivery and store purchase, thus lowering the online price: $p_{e} \leq c$.
The other maximizes demand from either BOPS or store purchase and sacrifices delivery demand, thus raising the online price: $p_{e} > c$.
In other words, in Regions BE-L and SE-L, demand comes from multiple consumer segments, while in Regions BE-U and SE-U, delivery demand is $0$.
When the subjective probability, $\theta$, is high, more consumers choose store purchase. Hence, the retailer can expect more demand from this option, thereby increasing the online price but sacrificing delivery demand.
If $\theta$ decreases, the retailer expects BOPS demand, sacrificing store purchase demand.
Therefore, the retailer’s profit depends on the subjective probability, $\theta$.
\par
Next, let us consider the profits that the retailer can maximize in each region and the respective prices.
Proposition \ref{prop:profit_max} shows the result of the profit maximization. (See Appendix \ref{sec:proof} for the proof.)
\begin{prop}
\label{prop:profit_max}
Table \ref{tbl:Solutions_F1_BOPS} shows the optimal prices and profits in Regions BE-L, BE-U, SE-L, SE-U, and S.
\end{prop}
\par
%
\begin{table}[h]
\caption{Optimal prices and profits}
\centering
\includegraphics[width=9.5cm,bb=0 0 297.75 420.75]{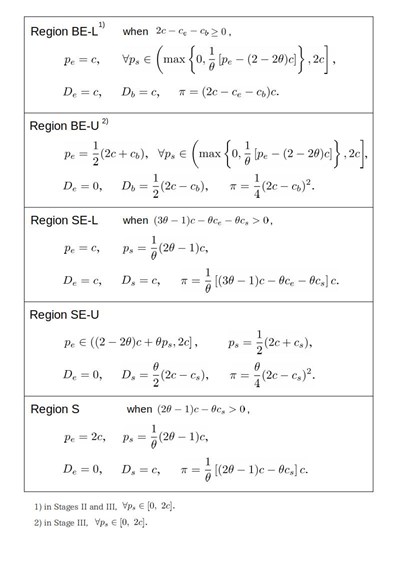}
\label{tbl:Solutions_F1_BOPS}
\end{table}
%
\noindent
Thus, we find corner solutions in Regions BE-L and SE-L, but interior solutions in Regions BE-U, SE-U, and S.
In Regions BE-L and SE-L with $p_{e} = c$, all consumers are equally divided into two segments; further, in both regions, the retailer obtains $c$ consumers who choose delivery.
Therefore, the retailer can profit from the two options in each region.
However, for optimal solutions to exist in these regions, the fulfillment costs for delivery and store purchase, $c_{e}$ and $c_{s}$, respectively, must be sufficiently low.
On the contrary, in Regions BE-U and SE-U with $p_{e} > c$, the retailer removes delivery from the two options.
Furthermore, it can obtain profits solely from BOPS in Region BE-U but solely from store purchase in Region SE-U, making it necessary to raise the store price, $p_{s}$, sufficiently.
\par
\bigskip
\noindent
\textbf{Case without offering BOPS} \nopagebreak
\par
\begin{figure}[htbp]
\centering
\includegraphics[width=8.5cm,bb=0 0 297.75 297.75]{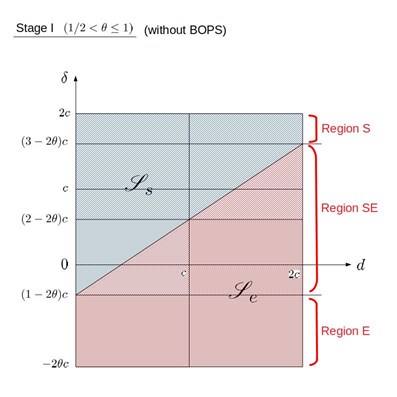}
\caption{Consumer segments (Stage I without BOPS)}
\label{fig:Segment_F1_NBOPS}
\end{figure}
\begin{figure}[htbp]
\centering
\includegraphics[width=8.5cm,bb=0 0 297.75 297.75]{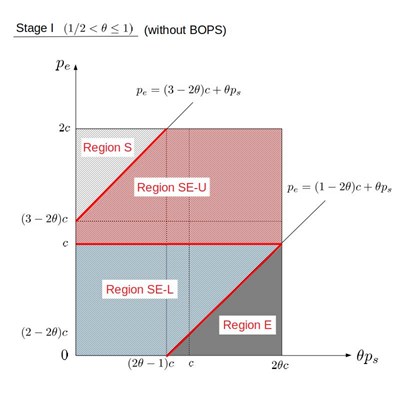}
\caption{Price regions (Stage I without BOPS)}
\label{fig:Region_F1_NBOPS}
\end{figure}

While the previous settings have assumed that the retailer offers three options, it can choose not to offer consumers BOPS.
Here, we consider the optimal profit when the retailer does not offer BOPS.
Figures \ref{fig:Segment_F1_NBOPS} and \ref{fig:Region_F1_NBOPS} show the consumer segment and price region, respectively, for Stage I only, when not offering BOPS.
\par
\vspace{\baselineskip}
\vspace{\baselineskip}
\noindent
Compared with the previous case, all consumers choose delivery in the region when $-2 \theta c \leq \delta \leq (1 - 2 \theta)c$, named \textbf{Region E}.
Thus, in this region, the retailer cannot sell in the physical store, and all sales are online.
The following proposition summarizes the result.
\begin{prop}
\label{prop:NBOPS}
If no BOPS option is offered, demand is $2c$ in Region E, where the optimal prices are $p_ {e} = c, \ p_ {s} = 2c$ and the maximum profit is $\pi = 2(c - c_{e})c$, which is non-negative as long as $c > c_ {e} $.
\end{prop}
Because $p_{e} \leq c$ in Region E, the retailer should clearly maximize its profit from all consumers with $p_{e} = c$.
In this way, it does not use the physical store in terms of either BOPS or store purchase but only sells online and delivers products from the fulfillment center.
Hereafter, we call this scenario \emph{online sales}.
\section{The retailer's optimal strategies}
\label{sec:strategies}
\noindent
\textbf{\textcolor{black}{The retailer's decision making}} \nopagebreak
\par
From the above results, we determine that the retailer can optimize its profit by adjusting the online and store prices to induce consumers toward profitable consumer segments.
Here, we discuss the optimal strategy of the retailer.
First, let us formulate its decision-making problem as follows:
\begin{description}
    \item[Step 0]\ \ \ The retailer observes the fulfillment costs, $c$, $c_{e}$, $c_{b}$, and $c_{s} $ and the subjective probability, $\theta$.
    \item[Step 1]\ \ \ It decides whether to offer the consumer the BOPS option.
    \item[Step 2]\ \ \ It forms the consumer segments.
    \item[Step 3]\ \ \ It sets the online and store prices to maximize its profit.
\end{description}
\noindent
Here, the decision making in Step 1 and Step 2 corresponds to choosing one of the five regions, BE-L, SE-L, BE-U, SE-U, S, or E.
Let us enumerate the possible strategies as follows.
\begin{description}
  \item[Strategy BEL:] Induce non-local consumers to choose delivery and local consumers to choose BOPS.
  \item[Strategy SEL:] Induce non-local consumers to choose delivery and local consumers to choose store purchase.
  \item[Strategy BEU:] Exclude non-local consumers from the market and induce some local consumers to choose BOPS.
  \item[Strategy SEU:] Exclude non-local consumers from the market and induce some local consumers to choose store purchase.
  \item[Strategy S:] Exclude all non-local consumers from the market and induce all local consumers to choose store purchase.
  \item[Strategy E:] Induce all consumers to choose delivery, with the physical store not used.
\end{description}
\noindent
From the previous discussion, we can determine that the maximum profit in each region optimized in Step 3 is as follows:
\begin{equation*}
\label{eq:profits}
\begin{aligned}
  & \piBEL \equiv  (2c - c_{e} - c_{b})c, \ \ \text{where}\ 2c - c_{e} - c_{b} \geq 0, \\
  & \piBEU \equiv  \dfrac{1}{4}(2c - c_{b})^{2}, \\
  & \piSEL \equiv  \dfrac{1}{\theta}\left[ (3\theta - 1)c - \theta c_{e} - \theta c_{s} \right]c, \\
  & \hspace{3cm} \text{where}\ (3\theta - 1)c - \theta c_{e} - \theta c_{s} \geq 0, \\
  & \piSEU \equiv  \dfrac{\theta}{4}(2c - c_{s})^{2}, \\
  & \piS \equiv  \dfrac{1}{\theta}\left[ (2\theta - 1)c - \theta c_{s} \right]c, \ \ \text{where}\ (2\theta - 1)c - \theta c_{s} \geq 0, \\
  & \piE \equiv 2(c - c_{e})c, \ \ \text{where}\ c > c_{e}.
\end{aligned}
\end{equation*}
\par
By comparing these profits, we can obtain the optimal strategy of the retailer.
However, some of these profits are not guaranteed to be non-negative, and it depends on the cost parameters.
Indeed, not all strategies are feasible if the shipping cost, $c$, is sufficiently low compared with the other costs.
Therefore, hereafter, we make the following assumption for all six feasible strategies above.
\begin{ass}
The fulfillment cost is always less than the shipping cost: $c_{e}, c_{b}, c_{s} < c$.
\end{ass}
By adopting this assumption, $\piBEL$ and $\piE$ are always non-negative and the existence of solutions in those regions is guaranteed.
Under this assumption, we define the following values:
\begin{equation*}
\label{eq:thetas}
\begin{aligned}
  & \thetaSEL \equiv \dfrac{c}{3c - c_{e} - c_{s}}, \ \ \ \ \
    \thetaBSL \equiv \dfrac{c}{c + c_{b} - c_{s}}, \\
  & \thetaBSU \equiv \dfrac{(2c - c_{b})^{2}}{(2c - c_{s})^{2}}, \ \ \ \ \
    \thetaSE \equiv \dfrac{c}{c + c_{e} - c_{s}}, \\
  & \thetaSn \equiv \dfrac{2c}{(2c - c_{s})^{2}} \\
  & \hspace{0.5cm} \times \left[ 3c - c_{e} - c_{s} - \sqrt{(c - c_{e})(5c - c_{e} - 2c_{s})} \right], \\
  & \thetaSp \equiv \dfrac{2c}{(2c - c_{s})^{2}} \\
  & \hspace{0.5cm} \times \left[ 3c - c_{e} - c_{s} + \sqrt{(c - c_{e})(5c - c_{e} - 2c_{s})} \right].
\end{aligned}
\end{equation*}
By comparing each profit, we have the following proposition.
\begin{prop}
\label{prop:comparison}
(1) $\piSEL > \piS$, $\piSEU \geq \piS$;
(2) If $\theta \geq \thetaSEL$, there exists a solution in Region SE-L;
(3) If $\theta \lessgtr \thetaBSL$, then $\piBEL \gtrless \piSEL$;
(4) If $\theta \lessgtr \thetaBSU$, then $\piBEU \gtrless \piSEU$;
(5) If $\theta \gtrless \thetaSE$, then $\piSEL \gtrless \piE$;
(6) If $c_{e} \gtrless c_{b}$, then $\piBEL \gtrless \piE$;
(7) If $4c^{2} - 4c_{e}c - c_{b}^{2} \gtrless 0$, then $\piBEL \gtrless \piBEU$;
(8) If $\thetaSn \leq \theta \leq \thetaSp$, then $\piSEL \geq \piSEU$. Otherwise, $\piSEL < \piSEU$.
\end{prop}
This proposition allows us to derive the following implications.
First, from (1), the retailer does not adopt Strategy S.
In this region, it must set the store price low to induce consumers to purchase at the store, and this is not profitable even if all local consumers purchase.
Next, when the sum of $c_{e}$ and $c_{s}$ is sufficiently high compared with $c$, Strategy SEL cannot be pursued unless the subjective probability $\theta$ is sufficiently high.
In the comparison between Strategy SEL and Strategy BEL, if $c_{s}$ is sufficiently high for $c_{b}$, Strategy SEL is not superior to Strategy BEL and vice versa.
Similarly, when $c_{s}$ is sufficiently high compared with $c_{b}$, Strategy SEU is not superior to Strategy BEU.
Comparing Strategy BEL and Strategy BEU, if $c_{e}$ is sufficiently high, Strategy BEU is adopted but not Strategy BEL.
Finally, if $c_{e}$ is sufficiently low compared with $c_{s}$ and $c_{b}$, then Strategy E is more likely to be adopted.
\par
\bigskip
\noindent
\textbf{Numerical examples} \nopagebreak
\par
To confirm the above proposition, we compare the profits in the following numerical examples in which $c = 1.0$.
\par
\bigskip
\noindent
\textbf{Case (1):} $c_{e} = 0.5$, $c_{b} = 0.4$, $c_{s} = 0.1$.
\par
\begin{figure}[htbp]
\centering
\includegraphics[width=5.5cm,bb=0 0 297.75 420.75]{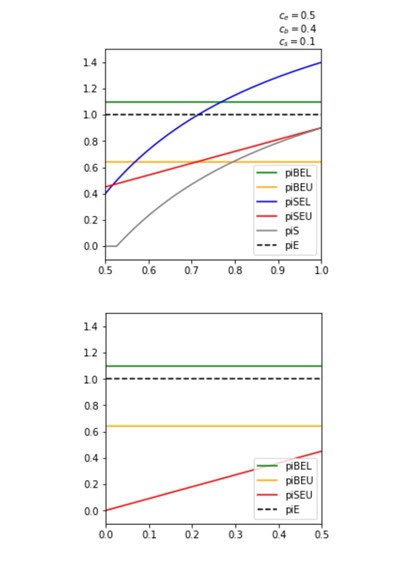}
\caption{Profits (Case (1): $c = 1.0, c_{e} = 0.5, c_{b} = 0.4, c_{s} = 0.1$)}
\label{fig:Profits_c10_ce05_cb04_cs01}
\end{figure}

\noindent
In this case in which $c_{e} > c_{b} > c_{s}$, we obtain the following numerical solutions: $\thetaSEL = 0.417$, $\thetaBSL = 0.769$, $\thetaBSU = 0.709$, $\thetaSE = 0.714$, and $\thetaSn = 0.517$.
Each profit in this case is shown in Figure \ref{fig:Profits_c10_ce05_cb04_cs01}.
\par
\vspace{\baselineskip}
\vspace{\baselineskip}
\noindent
In Stage I (above), when $\theta$ is close to $1$, the profit is higher in the order of $\piSEL$, $\piBEL$, and $\piE$.
However, when $\theta$ falls below $\thetaBSL = 0.769$, $\piBEL$ becomes higher than $\piSEL$, and in Stages \II and \III (below), $\piBEL$ is the highest profit ($\piSEL$ does not exist).
In all stages, $\piE$ does not exceed $\piBEL$.
In this case, if the subjective probability $\theta$ is sufficiently high, the retailer should adopt Strategy SEL and induce all local consumers to visit the store, while letting all non-local consumers choose delivery.
Conversely, if $\theta$ is low, it should pursue Strategy BEL, inducing all local consumers to collect the product ordered online at the store.
In this case, however, non-local consumers are not excluded from the market and all consumers can purchase.
In summary, in Case (1), the retailer's optimal strategy is
\begin{description}
    \item[Step 1]\ \ \ Offer consumers BOPS.
    \item[Step 2]\ \ \ If $\theta \geq \thetaBSL$, then pursue Strategy SEL and Strategy BEL otherwise.
    \item[Step 3]\ \ \ Set the optimal prices in each region according to Proposition \ref{prop:profit_max}.
\end{description}
\par
\bigskip
\noindent
\textbf{Case (2)} $c_{e} = 0.5$, $c_{b} = 0.6$, $c_{s} = 0.1$.
\par
\begin{figure}[htbp]
\centering
\includegraphics[width=5.5cm,bb=0 0 297.75 420.75]{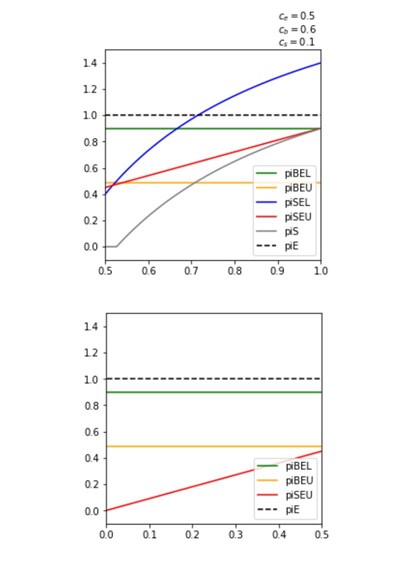}
\caption{Profits (Case (2): $c = 1.0, c_{e} = 0.5, c_{b} = 0.6, c_{s} = 0.1$)}
\label{fig:Profits_c10_ce05_cb06_cs01}
\end{figure}

\noindent
In this case in which $c_{b} > c_{e} > c_{s}$, we obtain the following numerical solutions: $\thetaSEL = 0.417$, $\thetaBSL = 0.667$, $\thetaBSU = 0.543$, $\thetaSE = 0.714$, and $\thetaSn = 0.517 $.
Each profit in this case is shown in Figure \ref{fig:Profits_c10_ce05_cb06_cs01}.
\par
\vspace{\baselineskip}
\vspace{\baselineskip}
\noindent
Unlike Case (1), in Stage I, when $\theta$ is close to $1$, the profit is higher in the order of $\piSEL$, $\piE$, and $\piBEL$.
When $\theta$ falls below $\thetaSE = 0.714$, $\piE$ becomes higher than $\piSEL$, and $\piE$ is the highest profit in Stages \II and \III.
In this case, $\piE$ dominates $\piBEL$ in all stages.
In other words, in this case, if the subjective probability $\theta$ is sufficiently high, the retailer adopts Strategy SEL.
Even if $\theta$ is low, the retailer does not pursue Strategy BEL but rather adopts Strategy E with the physical store no longer used.
In summary, in Case (2), the retailer's optimal strategy is
\begin{description}
    \item[Step 1]\ \ \ Do not offer BOPS.
    \item[Step 2]\ \ \ If $\theta \geq \thetaSE$, adopt Strategy SEL and Strategy E otherwise.
    \item[Step 3]\ \ \ Set the optimal prices in each region according to Propositions \ref{prop:profit_max} and \ref{prop:NBOPS}.
\end{description}
\par
\bigskip
\noindent
\textbf{Case (3):} $c_{e} = 0.9$, $c_{b} = 0.8$, $c_{s} = 0.7$.
\par
\begin{figure}[htbp]
\centering
\includegraphics[width=5.0cm,bb=0 0 297.75 420.75]{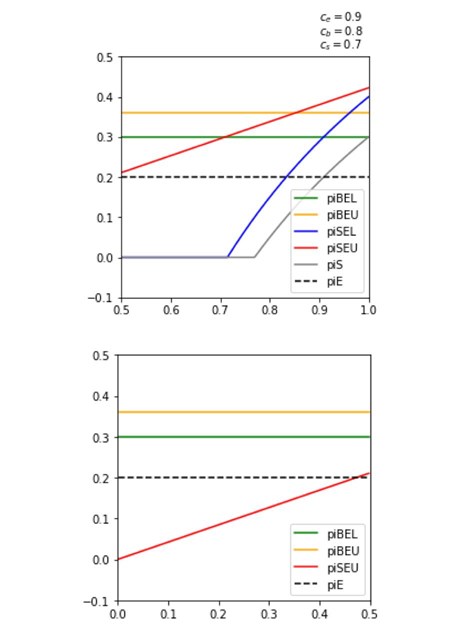}
\caption{Profits (Case (3): $c = 1.0, c_{e} = 0.9, c_{b} = 0.8, c_{s} = 0.7$)}
\label{fig:Profits_c10_ce09_cb08_cs07}
\end{figure}

In this case, $c_{b}$, $c_{e}$, and $c_{s}$ are sufficiently high compared with $c = 1.0$, $\piBEU > \piBEL $, and $\piSEU > \piSEL$.
The numerical solutions are $\thetaSEL = 0.714$, $\thetaBSL = 0.909$, $\thetaBSU = 0.852$, $\thetaSE = 0.833$, and $\thetaSn = 0.104$.
Each profit in this case is shown in Figure \ref{fig:Profits_c10_ce09_cb08_cs07}.
\par
\vspace{4\baselineskip}
\vspace{\baselineskip}
\noindent
As shown in this figure, in Stage I, $\piSEU$ is the highest when $\theta$ is close to $1$, but $\piBEU$ is higher when $\theta$ is less than $\thetaBSU = 0.852$.
It is higher than $\piSEU$, and $\piBEU$ has the highest profit in Stages \II and \III.
In this case, $\piE$ does not exceed $\piBEL$ and $\piBEU$ in all stages.
In other words, if the subjective probability $\theta$ is sufficiently high, the retailer adopts Strategy SEU; on the contrary, if $\theta$ is low, it adopts Strategy BEU.
In summary, in Case (3), the retailer's optimal strategy is
\begin{description}
    \item[Step 1]\ \ \ Offer consumers BOPS
    \item[Step 2]\ \ \ If $\theta \geq \thetaBSU$, then pursue Strategy SEU and Strategy BEU otherwise.
    \item[Step 3]\ \ \ Set the optimal prices in each region according to Proposition \ref{prop:profit_max}.
\end{description}
\par
\begin{figure}[htbp]
\includegraphics[width=8.0cm,bb=0 0 297.75 297.75]{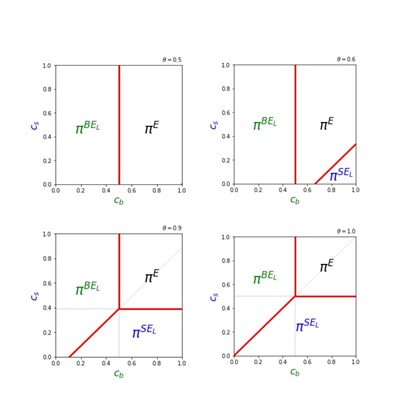}
\caption{Cost regions where BOPS, store purchase, and online sales are optimal ($c_{e} = 0.5$)}
\label{fig:Cost_theta}
\end{figure}

\bigskip
In the above numerical examples, we compare the profits when the subjective probability changes, with the fulfillment cost fixed.
Conversely, we can consider how the optimal strategy varies under a certain subjective probability depending on the fulfillment cost.
Figure \ref{fig:Cost_theta} shows the optimal strategies when $c_{e} = 0.5$.
\par
%
\vspace{\baselineskip}
\noindent
This figure describes the regions in which Strategy BEL, Strategy SEL, and Strategy E are optimal on the $(c_{b}\!-\!c_{s})$ plane.
As shown, when the subjective probability $\theta$ is high, if $c_{s}$ is sufficiently low, the region size in which Strategy SEL is optimal is large.
Conversely, if $\theta$ falls, the region size shrinks and disappears when $\theta = 0.5$.
\section{Simple dynamic simulations}
\label{sec:simulation}
\noindent
Finally, we perform simple dynamic simulations to consider the sales strategy the retailer should adopt when it sells the product in a finite time horizon. Two types of simulations are examined.
\par
\begin{figure}[htbp]
\centering
\includegraphics[width=6.0cm,bb=0 0 297.75 420.75]{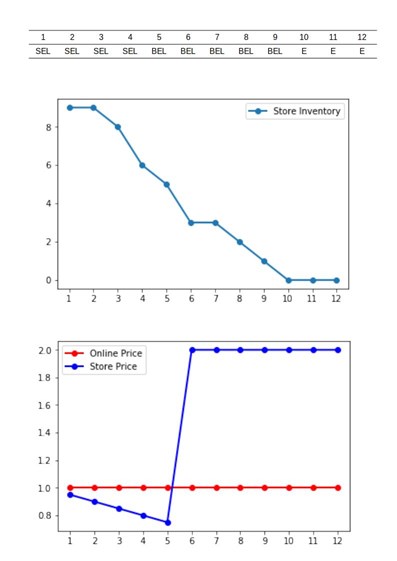}
\caption{The first simulation result (path of optimal strategies, inventory, and prices)}
\label{fig:Simulation_alpha005}
\end{figure}

In the first simulation, consider a seasonal product sold over 12 periods (hereafter, \emph{season}).
The retailer sets the online and store prices for each period after knowing the subjective probability at the beginning of the period.
The subjective probability decreases over time.
This probability transition is not correlated with the actual inventory amount but changes by a specific function for simplification.
Here, let the subjective probability in period $t$ be $\theta_{t} = \frac{1}{1 + \alpha t}$.
The fulfillment costs are the same as in Case (1) above.
We assume that the inventory is determined at the beginning of the season and not replenished.
Further, the inventory at the fulfillment center is sufficiently large and will not be exhausted, while the physical store's inventory amount is $10.0$.
Therefore, if the stock runs out within the season, the retailer must take only online sales after that.
Furthermore, to make the model include uncertainty, assume that the demand of the physical store in each period is a random variable $X$ that follows the Poisson distribution, $P(X = k; c) = \frac{e^{-c}c^{k}}{k!}, \ (K = 0, 1, 2, \ldots)$.
Figure \ref{fig:Simulation_alpha005} presents the simulation result, showing  that Strategy SEL is adopted from period $1$ to $4$ and Strategy BEL from period $5$ to $9$.
\par
%
\vspace{\baselineskip}
\noindent
Because the physical store's inventory is exhausted in period $10$, Strategy E is adopted with online sales only after this period.
We refer to the period in which the retailer changes its strategy as the \emph{switching period}.
\par
The second simulation repeats the first simulation 1000 times, while the subjective probability is updated to be a 95\% probability of changing.
Table \ref{tbl:Simulation2} shows the results for three scenarios: the retailer always offers BOPS during the season (Scenario 1), does not offer it (Scenario 2), and changes the BOPS strategy following our optimal strategy (Scenario 3).
\par
%
\begin{table}[h]
\caption{The second simulation result}
\par
\vspace{1cm}
\centering
\includegraphics[width=8.5cm,bb=0 0 396.75 222.75]{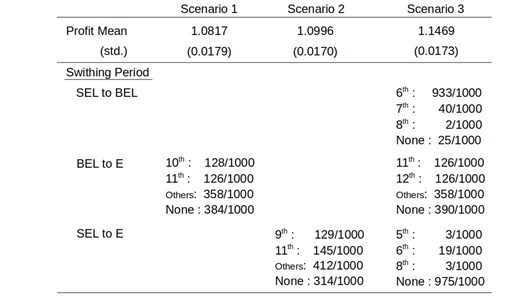} \label{tbl:Simulation2}
\end{table}
\vspace{\baselineskip}
\noindent
The simulation results show that Scenario 3 increases the retailer's profit by 6\% compared with Scenario 1 and 4.3\% compared with Scenario 2 on average.
In Scenario 3, of the 1000 sample paths, the retailer switches from Strategy SEL to Strategy BEL in period 6 for 933 paths and does not switch for 25 paths (referred to as NONE).
Of these 25 paths, it moves to Strategy E in period 6 for 19 paths.
\par
In summary, the retailer offers the BOPS option in later as opposed to in earlier periods.
We also find the same implication with the cost parameters in Cases (2) and (3).
\section{Conclusion}
\label{sec:conclusion}
This study analyzed how a retailer can adjust its prices and segmented consumers to solve profit optimization problems with fulfillment costs.
We found that the retailer must form consumer segments according to their subjective probability and select the optimal strategy by setting appropriate prices.
\par
While retailers often seem indifferent to BOPS timing in the real-world distribution setting, this study concludes that BOPS is a robust strategy that raises retailers' profits.
Indeed, when the subjective probability decreases over the season, retailers must switch their BOPS strategy at some point to maximize their profit. Specifically, they should offer BOPS in later periods rather than in earlier ones.
As shown by our simulation results, neither (1) always offering BOPS during the season nor (2) never offering BOPS is optimal.
Retailers can increase their profit by 6\% compared with always offering BOPS and 4.3\% compared with never doing so by switching the BOPS strategy at the optimal timing.
Hence, one managerial implication of our findings is that retailers should use data analysis to correctly estimate consumers' subjective probability to determine when to switch their BOPS strategy.
\par
Finally, we discuss the limitations of our model because it relies strongly on certain simplifying assumptions.
First, we assumed that when consumers choose to visit the physical store directly, they do not purchase if the product is out of stock.
However, in reality, consumers may try to purchase online instead.
While such an analysis would be more complicated, the motivation for choosing store purchase may be lower than it is in our model.
\par
Second, we assumed that consumers' subjective probability is exogenous and uncorrelated with actual inventory.
This setting contrasts with the rational expectations assumption of \citet{Gao_Su2016} and \citet{Kusuda2019} and ignores the information available for consumers.
In the real world, retailers can provide consumers with their inventory status online and consumers can form their subjective probability based on such information.
The analysis of optimal prices under such endogenous subjective probabilities thus remains an issue for future research.
\par
Third, there was only one physical store in this model and consumers only used the Internet or the physical store.
However, real-world distribution systems use a network of physical stores.
Analyzing the seamless connection between multiple physical stores is therefore another remaining issue for omnichannel analyses.
\par
Finally, while we performed simple dynamic simulations, it is strictly a ``myopic'' model because the retailer determines the current prices and sales strategy based on the subjective probability of that period in a pointwise manner.
Analyzing a ``full'' dynamic model in which the retailer determines the price paths throughout the season using a dynamic programming method is also an issue for the future.
\bibliography{kusuda_JPIF_arXiv}
\bibliographystyle{apa}
\appendix
\section{Proofs of the propositions}
\label{sec:proof}
\par
\bigskip
\noindent
\textbf{Proof of Proposition \ref{prop:demand}} \nopagebreak
\par
Region BE-L: From the definition of the region, it is $p_{e} \leq c$, which means $u_{e} \geq 0$ for all consumers; therefore, $D_{e} = c$.
Next, denote the length of the interval of segment $\mathscr{S}_{i}$ by $|\mathscr{S}_{i}|$ ($i = e, b, s$),
and $d_{b}$ such that $u_{b} = 0$ by $\bar{d}_{b}$.
Then, we have the following relationship:
\begin{equation*}
  |\mathscr{S}_{b}| - \bar{d}_{b} = c - (2c - p_{e}) \gtrless 0 \ \Longleftrightarrow \ p_{e} \gtrless c.
\end{equation*}
In this region, as $p_{e} \leq c$, $|\mathscr{S}_{b}| \leq \bar{d}_{b}$, which means $D_{b} \equiv \min \{ |\mathscr{S}_{b}|, \ \bar{d}_{b} \} = c$.
\par
Region BE-U: $D_{e} = 0$ because $p_{e} > c$.
In addition, as $|\mathscr{S}_{b}| > \bar{d}_{b}$, $D_{b} \equiv \min \{ |\mathscr{S}_{b}|, \ \bar{d}_{b} \} = 2c - p_{e}$.
\par
Region SE-L: Because $p_{e} \leq c$ from the definition of the region, $D_{e} = c$.
Denoting $d_{s}$ such that $ u_{s} = 0$ by $\bar{d}_{s}$, we have the following relationship:
\begin{equation*}
\begin{aligned}
  & |\mathscr{S}_{s}| - \bar{d}_{s} = \left( \delta -(1 - 2\theta)c \right) - (2\theta c - \theta p_{s}) \gtrless 0 \\
  & \Longleftrightarrow \ p_{e} \gtrless c.
\end{aligned}
\end{equation*}
In this region, as $p_{e} \leq c$, $D_{s} \equiv \min\{ |\mathscr{S}_{s}|, \ \bar{d}_{s} \} = \delta - (1 - 2\theta)c$.
\par
Region SE-U: As $p_{e} > c$, $D_{e} = 0$.
In addition, $|\mathscr{S}_{s}| > \bar{d}_{s}$; hence, $D_{b} = 2\theta c - \theta p_{s}$.
\par
Region S: Similar to the above, we have $D_{s} = 2\theta c - \theta p_{s}$.
$\blacksquare$
\par
\bigskip
\noindent
\textbf{Proof of Proposition \ref{prop:profit_max}} \nopagebreak
\par
Region BE-L:
The retailer sells to consumers in segment $\mathscr{S}_{e}$ (demand $c$) at price $p_{e}$, which costs $c_{e}$ per unit.
Therefore, it obtains profit $(p_{e} - c_{e})c$ in this segment.
Similarly, the retailer obtains profit $(p_{e} - c_{b})c$ from the consumers in segment $\mathscr{S}_{b}$; hence, total profit is $\pi = (p_{e} - c_{e})c + (p_{e} - c_{b})c$.
This profit is maximized with respect to $p_{e} \leq c$, which determines the optimal price, $p_{e} = c$, and the profit, $\pi = (2c - c_{e} - c_{b})c$, unless $2c - c_ {e} - c_{b} \geq 0$.
\par
Region BE-U:
Because $p_{e} > c$, $D_{e} = 0$, and the profit can be obtained only from consumers in segment $\mathscr{S}_{b}$.
Thus, the retailer maximizes $\pi = (p_{e} - c_{b})(2c - p_{e})$ with respect to $p_{e}$.
Then, there is an interior solution in $(c, \ 2c]$; therefore, $p_{e} = \frac{1}{2}(2c + c_{b})$.
The profit is $\pi = \frac{1}{4}(2c - c_{b})^{2}$.
\par
Region SE-L:
From Proposition \ref{prop:demand}, the profit to be maximized is
\begin{equation*}
\begin{aligned}
  & \pi = (p_{e} - c_{e})\left[ (3 - 2\theta)c - p_{e} + \theta p_{s} \right] \\
  & + (p_{s} - c_{s})\left[ -(1 - 2\theta)c - \theta p_{s} + p_{e} \right].
\end{aligned}
\end{equation*}
From $(p_{e}-\theta p_{s})$ in Figure \ref{fig:Region_F1_BOPS},
the slope of the iso-profit curve at profit level $\bar{\pi}$ is
\begin{equation*}
  \left. \frac{dp_{e}}{d\theta p_{s}} \right|_{\bar{\pi}}
  = - \frac{(3 - 2\theta)c + c_{e} - c_{s} - 2p_{e} + (1 + \theta)p_{s}}{-(1 - 2\theta)c - \theta c_{e} + \theta c_{s} - 2\theta p_{s} + (1 + \theta)p_{e}} \cdot \frac{1}{\theta}.
\end{equation*}
However, if $p_{s}$, whose slope is $0$ when $p_{e} = c$, is $p_{s}^{\ast}$, then $\theta p_{s}^{\ast} > (2\theta - 1)c$; hence, the solution in Region SE-L is the corner solution: $(p_{e}, \ \theta p_{s}) = (c, \ (2\theta - 1)c)$ (point A in Figure \ref{fig:Region_F1_BOPS}).
If $(3\theta - 1)c - \theta c_{e} - \theta c_{s} \geq 0$, the profit is non-negative, meaning that this price is the optimal solution and the profit is $\pi = \frac{1}{\theta} \left[ (3\theta - 1)c - \theta c_{e} - \theta c_{s} \right] c$.
\par
Region SE-U:
Because $D_{e} = 0$,
the retailer only earns the profit from segment $\mathscr{S}_{b}$ and maximizes $\pi = (p_{e} - c_{b})(2\theta c - \theta p_{s})$.
Then, there is an interior solution in $\left( \frac{1}{\theta}(2\theta - 1)c, \ 2c \right]$; thus, $p_{s} = \frac{1}{2}(2c + c_{s})$.
The profit is $\pi = \frac{\theta}{4}(2c - c_{s})^{2}$.
\par
Region S:
Because there are no interior solutions, the corner solution $(p_{e}, \ \theta p_{s}) = (2c, \ (2\theta - 1)c)$ is optimal (point B in Figure \ref{fig:Region_F1_BOPS}), which is non-negative if $(2\theta - 1)c - \theta c_{s} \geq 0$.
The profit is $\pi = \frac{1}{\theta}\left[ (2\theta - 1)c - \theta c_{s} \right]c$.
$\blacksquare$
\end{document}